\documentclass[prd,twocolumn]{revtex4}\begin{document}\preprint{}
\title{An inhomogeneous solution of Einstein equations with  viscous   fluids }
\author{ Z. Haba\\
Institute of Theoretical Physics, University of Wroclaw,\\ 50-204
Wroclaw, Plac Maxa Borna 9, Poland}
\email{zbigniew.haba@uwr.edu.pl}
\begin{abstract}Assuming conformally flat metric we obtain inhomogeneous solutions of  Einstein equations
with the energy-momentum of a viscous fluid. We suggest that the
viscous solution can be applied as a model of an expanding
inhomogeneous dark energy.
\end{abstract}
\maketitle
\section{Introduction}Einstein equations are studied with various
assumptions on the energy-momentum of "matter". The quantum form
of the equations with the energy-momentum of the fields of the
standard model cannot be realized because of the difficulties with
the quantization of the gravitational field. We must use
approximations for the rhs of Einstein equations. The most studied
models involve a perfect fluid on the rhs of Einstein equations.
The solutions can describe an expanding universe or compact
objects like the static as well as expanding and collapsing stars
\cite{gravitation}\cite{weinberg} (for  inhomogeneous solutions
and their relevance in gravity see
\cite{solutions}\cite{bolejko}\cite{flan}\cite{mukh}\cite{ellis}).
However, it is obvious that the assumption of a perfect fluid on
the rhs of Einstein equations, usually applied in gravitational
models, is an idealization. All physical fluids at high
temperature have a non-zero viscosity. It is not simple to include
viscosity in the solutions of Einstein equations. Concerning the
homogeneous solutions, it is known that the FLWR form of the
solution  admits only the bulk viscosity. The effect of the bulk
viscosity on the expansion of the universe has been discussed in
\cite{weinberg}\cite{klimek} \cite{odintsov}\cite{odintsov2}. The
introduction of the bulk viscosity modifies the formula for the
pressure which leads to interesting reformulations in the dynamics
and thermodynamics depending on the change of the equation of
state \cite{odintsov}\cite{odintsov2}\cite{brevik0}. There are
many examples of inhomogeneous solutions of Einstein equations
with a perfect fluid on the rhs \cite{solutions}\cite{bolejko}.
However, it seems that no examples are known of inhomogeneous
solutions with a shear viscosity of the fluid. Recently the
relativistic fluid equations have been studied in the heavy-ion
physics (with a suggestion of a possible simulation of the Big
Bang, see the review \cite{hunperfect}). Some solutions of the
hydrodynamics equations for the perfect fluid have been obtained,
(see ,e.g., \cite{hunperfect2}). Little is known about solutions
with viscosity (see however \cite{hungarian}). Another approach to
such models is developed in refs.\cite{ion1}\cite{ion2}\cite{ion3}
where the Einstein tensor for various metrics is interpreted as an
energy-momentum tensor of a fluid (including possibly a viscous
fluid).

 In this paper we assume that the metric is
conformally flat. A large class of models can be expressed in a
conformally flat form including  the FLWR solutions \cite{infeld}
\cite{taubes}. However, the FLWR metric admits only bulk viscosity
of the fluid. We obtain conformally flat inhomogeneous solutions
with a non-zero shear and bulk viscosity which satisfies the dark
energy equation of state
 $\rho=-p$. It is known that if the equation of state for the
energy-momentum of a perfect fluid has the form  $\rho=-p$ then
from the conservation law it follows that $\rho=const$. In such a
case the dark energy is just the cosmological constant. We obtain
a solution in the form of an expanding    viscous fluid. The
expansion is at lower rate than the expansion of radiation and of
the dust. The resulting viscous fluid could be a candidate for a
dark energy dominating the energy-momentum at large time.

\section{The energy-momentum tensor}
We consider Einstein equations
\begin{equation}
G_{\mu\nu}=8\pi G T_{\mu\nu},
\end{equation}
where $G_{\mu\nu}=R_{\mu\nu}-\frac{1}{2}g_{\mu\nu}R$ is the
Einstein tensor and $T_{\mu\nu}$ is the energy-momentum. We set
the velocity of light c=1 and the length will be measured in
Planck units $\sqrt{8\pi G}$ (so coordinates will be
dimensionless), hence we set $\sqrt{8\pi G}=1$ from now on. In the
case of a dust with a density $\rho$  the energy-momentum
$T_{\mu\nu}=\rho v_{\mu}v_{\nu}$ is conserved
$\nabla_{\mu}T^{\mu\nu}=0$ (together with the current
$\nabla_{\mu}(\rho v^{\mu})=0$),  where the relativistic velocity
$v_{\mu}$ satisfies the equation
\begin{equation} g^{\mu\nu}v_{\mu}v_{\nu}=1.
\end{equation} We can apply the Hamilton-Jacobi theory to express
the velocity by the action S, $v_{\mu}=\frac{\partial S}{\partial
x^{\mu}}$. Then, equation (2) reads
\begin{equation}
g^{\mu\nu}\frac{\partial S}{\partial x^{\mu}}\frac{\partial
S}{\partial x^{\nu}}=1
\end{equation}
Eqs.(1) and (3) can be considered as a system of equations which
determine the metric $g_{\mu\nu}$ and the fluid velocities
$v_{\mu}$.
  We extend this scheme to the viscous energy-momentum
\begin{equation}\begin{array}{l}T_{\mu\nu}=(\rho+p)v_{\mu}v_{\nu}
-g_{\mu\nu}p-\eta_{g}
(\nabla_{\mu}v_{\nu}+\nabla_{\nu}v_{\mu})\cr-\gamma_{g}
g_{\mu\nu}g^{\alpha\beta}\nabla_{\alpha}v_{\beta},
\end{array}\end{equation}where $p$ is the pressure.
In physical fluids the shear $\eta$ and bulk $\gamma$ viscosities
depend in an involved way on the temperature and density of the
fluids. In our solution the viscosities depend on the metric
through its scale factor (the viscosities are decreasing with an
expansion of the fluid).

 We use the decomposition
\begin{displaymath}
g^{\alpha\beta}=v^{\alpha}v^{\beta}+H^{\alpha\beta}
\end{displaymath}
where
\begin{displaymath}
H^{\alpha\beta}=g^{\alpha\beta}-v^{\alpha}v^{\beta}
\end{displaymath} in order to rewrite the energy-momentum (4) in
the standard form of Landau-Lifshitz \cite{LL} and Weinberg
\cite{weinberg}\cite{weinberg2}
\begin{equation}\begin{array}{l}
T^{\mu\nu} =(p_{e}+\rho_{e})v^{\mu}v^{\nu}-p_{e}g^{\mu\nu}\cr
-\eta_{g}
H^{\mu\alpha}H^{\nu\beta}(\nabla_{\alpha}v_{\beta}+\nabla_{\beta}v_{\alpha}
-\frac{2}{3}g_{\alpha\beta}g^{\sigma\lambda}\nabla_{\sigma}v_{\lambda})\cr
-\gamma_{g}H^{\mu\nu}g^{\alpha\beta}\nabla_{\alpha}v_{\beta}-
\kappa_{g}
(H^{\mu\lambda}v^{\nu}+H^{\nu\lambda}v^{\mu})Q_{\lambda},
\end{array}\end{equation}
where the heat current $Q$ is
\begin{equation}
Q_{\lambda}=\partial_{\lambda}T_{g}+T_{g}v^{\alpha}\nabla_{\alpha}v_{\lambda}
 \end{equation}
with the heat conductivity $\kappa_{g}$ and temperature $T_{g}$
satisfying the relations $\partial_{\lambda}T_{g}=0$ and
$\kappa_{g} T_{g}= \eta_{g}$. So,
\begin{displaymath}
\kappa_{g} Q_{\lambda}=\eta_{g}
v^{\alpha}\nabla_{\alpha}v_{\lambda}.
 \end{displaymath}
The effective pressure and density are
\begin{equation}
p_{e}=p-\frac{2}{3}\eta_{g}
g^{\sigma\lambda}\nabla_{\sigma}v_{\lambda},
\end{equation}\begin{equation}
\rho_{e}=\rho-\gamma_{g}
g^{\sigma\lambda}\nabla_{\sigma}v_{\lambda}.
\end{equation}
We  shall require for our solution in sec.4 that $\rho+p=0$ then
\begin{equation}
\rho_{e}+p_{e}=-(\frac{2}{3}\eta_{g}+\gamma_{g})
g^{\sigma\lambda}\nabla_{\sigma}v_{\lambda}.
\end{equation}
As is well-known \cite{LL}\cite{weinberg} the form (5) of the
energy-momentum with positive $\eta, \kappa$ and $\gamma$ ensures
the increase of the entropy.

 If eqs.(1) are to be non-contradictory we must have
\begin{equation}g^{\alpha\mu}\nabla_{\alpha}T_{\mu\nu}=0.
\end{equation}
Eqs.(10) can be considered as differential equations (relativistic
Navier-Stokes equations) relating $\rho,p$ and $v$. Eqs.(10) with
the metric $g_{\mu\nu}$ solving Einstein equations (1) follow from
the Einstein equations. However, when $v_{\alpha}$ satisfy (10) on
a certain manifold with the metrics $g_{\mu\nu}$ then only a
subset of $(g_{\mu\nu},v_{\alpha})$ will satisfy both eq.(10) and
eq.(1).
\section{The conformally flat metric}

We consider the conformally flat metric in four space-time
dimensions
\begin{equation}
ds^{2}=a(x)^{2}(d\tau^{2}-d{\bf
x}^{2})=a(x)^{2}\eta^{\mu\nu}dx_{\mu}dx_{\nu},
\end{equation}where $\eta^{\mu\nu}$  is the Minkowski metric.
 Then, the Einstein tensor
 can be expressed in the form
 \cite{blaschke}\cite{solutions}\cite{ijgm}

\begin{equation}
G_{\mu\nu}=(\tilde{\rho}+\tilde{p})\tilde{u}_{\mu}\tilde{u}_{\nu}-\tilde{p}g_{\mu\nu}+\Pi_{\mu\nu},
\end{equation}
where the fluid velocity is defined by
\begin{equation}
\tilde{u}_{\mu}=\partial_{\mu}a
\Big(g^{\mu\nu}\partial_{\mu}a\partial_{\nu}a\Big)^{-\frac{1}{2}}.
\end{equation}
The energy density is
\begin{equation}
\tilde{\rho}=3a^{-2}g^{\mu\nu}\partial_{\mu}a\partial_{\nu}a-g^{\mu\nu}\Pi_{\mu\nu}
\end{equation}
the pressure
\begin{equation}
\tilde{p}=a^{-2}g^{\mu\nu}\partial_{\mu}a\partial_{\nu}a+g^{\mu\nu}\Pi_{\mu\nu},
\end{equation}
where
\begin{equation}
\Pi_{\mu\nu}=-2a^{-1}\partial_{\mu}\partial_{\nu}a.
\end{equation}

Inserting the velocities (13) in eq.(12) we can also express
$G_{\mu\nu}$ in the form
\begin{equation}
G_{\mu\nu}=4a^{-2}\partial_{\mu}a\partial_{\nu}a-\tilde{p}g_{\mu\nu}+\Pi_{\mu\nu},
\end{equation}
Eqs.(12)-(16) suggest that $\tilde{u}_{\mu}\simeq \partial_{\mu}a$
can be a solution of Einstein equations (1). In the next section
we show that this is really the case.
\section{Einstein equations with a viscous fluid }
In eq.(4) the covariant derivative is expressed by the Christoffel
connection in conformally flat space \cite{blaschke}(from now on
the indices will be raised by means of the Minkowski metric)
\begin{equation}
\Gamma^{\lambda}_{\mu\nu}=\delta^{\lambda}_{\mu}\partial_{\nu}\ln
a+\delta^{\lambda}_{\nu}\partial_{\mu}\ln a
-\eta_{\mu\nu}\partial^{\lambda}\ln a
\end{equation}

 It is easy to see that the equation
\begin{equation}
G_{0j}=T_{0j}=-\eta_{g}(\nabla_{j}v_{0}+\nabla_{0}v_{j})
\end{equation}
is satisfied if  the equation of state is
\begin{equation}
\rho+p=0
\end{equation}the velocity
\begin{equation}
v_{\mu}=\frac{\partial}{\partial x^{\mu}}a
\end{equation} and\begin{equation} \eta_{g}=a^{-1}.
\end{equation}
However, if the solution (21) of eq.(19) is to be the relativistic
 velocity then the normalization (2) must be satisfied i.e.
$g^{\mu\nu}\partial_{\mu}a\partial_{\nu}a=1$. We write
$a=\exp(\psi)$. Then, on the basis of eqs.(2) and (21) $\psi$
satisfies the Hamilton-Jacobi equation
\begin{equation}
\eta^{\mu\nu}\partial_{\mu}\psi\partial_{\nu}\psi=1.
\end{equation}
This is the Hamilton-Jacobi equation (as described at eq.(3)) for
a free particle (with the unit mass) in the Minkowski space (its
velocity is $v_{\mu}=\partial_{\mu}\psi$).

The remaining components of the energy-momentum tensor are
\begin{equation} T_{00}=-pa^{2}-2 a^{-1}\nabla_{0}v_{0}-\gamma_{g}
(\nabla_{0}v_{0}-\nabla_{k}v_{k}),
\end{equation}where

\begin{equation}\begin{array}{l}
\nabla_{0}v_{0}-\nabla_{k}v_{k}\cr=\partial_{0}v_{0}-\partial_{k}v_{k}+2a^{-1}v_{0}\partial_{0}a
-2a^{-1}v_{k}\partial_{k}a\end{array}
\end{equation} and
\begin{equation}\begin{array}{l}
T_{jk}=\delta_{jk}\Big(pa^{2} -2(a^{-2}\partial_{r}a
v_{r}-a^{-2}\partial_{0}av_{0})\cr -\gamma_{g}
(\partial_{0}v_{0}-\partial_{j}v_{j}+2a^{-1}v_{0}a-2\partial_{r}v_{r})\Big)
\cr
-a^{-1}\Big(\partial_{j}v_{k}+\partial_{k}v_{j}-2a^{-1}\partial_{j}av_{k}
-2a^{-1}\partial_{k}av_{j}\Big).
\end{array}
\end{equation}
It can be seen that the non-diagonal parts of $G_{jk}$ (17) and
$T_{jk}$ (26) coincide if  the velocity  is determined by eq.(21)
and $\eta_{g}=a^{-1}$ (eq.(22)). There remain the diagonal
($\delta_{jk}$) parts of eqs.(26) and (17). They are equal if
\begin{equation}
p=(-3(a\gamma_{g}+1)-(a\gamma_{g}+2)\partial^{\mu}\partial_{\mu}\psi)\exp(-2\psi).
\end{equation}
The 00-equation  as derived from eqs.(24) and (17) is satisfied if
\begin{equation}
-p=(1+2a\gamma_{g})a^{-2}\eta^{\mu\nu}\partial_{\nu}a\partial_{\mu}a
+(2+a\gamma_{g})a^{-1}\eta^{\mu\nu}\partial_{\nu}\partial_{\mu}a
\end{equation}
It coincides with eq.(27) (which followed from the spatial
diagonal part of eq.(4)). So Einstein equations determine $p$ and
$v_{\mu}$ in terms of $\psi$ satisfying the Hamilton-Jacobi
equation (23). The relativistic Navier-Stokes equations for
$v_{\mu}$ follow from eq.(10). These equations can be obtained
simply by differentiation of eq.(28) and an insertion of
$\partial_{\mu}a=v_{\mu}$ because such a velocity is the solution
of these equations.

The energy and pressure are functions of the solution $\psi$ of
eq.(23). A construction of the general solution in terms of
characteristics is discussed in mathematical literature \cite{HJ}.
The solution can also be obtained by a calculation of the action
for  the relativistic Hamiltonian $\sqrt{1+{\bf v}^{2}}$
\cite{arnold}. Let us consider some special cases.
$\psi(x)=q_{\mu}x^{\mu}$  with $q_{\mu}q^{\mu}=1$ is the solution
of eq.(23). The next example is $\psi(x)=\sqrt{x^{2}}$ where
$x^{2}=x_{\mu}x^{\mu}$. It is the solution of eq.(23) for
time-like $x$. However, let us note that in these cases the
viscosity term in eq.(17) (as well as eq.(2))can be written in the
form
\begin{displaymath}
\Pi_{\mu\nu}=(\delta\rho+\delta p)v_{\mu}v_{\nu}-\delta p
g_{\mu\nu}
\end{displaymath}with certain $\delta\rho$ and $\delta p$.
Hence, the viscosity terms only modify the definition of $\rho$
and $p$. In the first case
$\nabla_{\mu}v_{\nu}+\nabla_{\nu}v_{\mu}\simeq
q_{\mu}q_{\nu}\simeq v_{\mu}v_{\nu}$ (the perfect fluids of
ref.\cite{ijgm}). In the second case
$\nabla_{\mu}v_{\nu}+\nabla_{\nu}v_{\mu}\simeq v_{\mu}v_{\nu}
+\eta_{\mu\nu}\delta p$ .

A non-trivial viscosity results from the solution
\begin{displaymath}
\psi=\sqrt{C_{2}-\tau-r}\sqrt{C_{1}-\tau+r},
\end{displaymath}
where $r=\vert {\bf x}\vert$. It solves eq.(23) if $\tau<C_{1}+r$
and $\tau<C_{2}-r$. We can get a solution for a large time
$\tau>C_{1}+r$ and $\tau>-C_{2}+r$ if we write $\psi$ in the
form\begin{equation}\psi=\sqrt{-C_{2}+\tau+r}\sqrt{-C_{1}+\tau-r}.
\end{equation}If $C_{1}=C_{2}=0$ then we return to the solution
$\sqrt{x^{2}}$.

We obtain a local   solution of eq.(23) depending on all variables
if we treat eq.(23) as the Hamilton-Jacobi equation for the
Hamiltonian $\sqrt{1+{\bf v}^{2}}$. Through the separation of
variables in cylindrical coordinates $(r,\phi,z)$ we obtain
\begin{equation} \psi=-E\tau+L\phi+f(r)+\gamma z
\end{equation}with the constant angular velocity $v_{\phi}=L$
and radial velocity
\begin{displaymath}
v_{r}=\frac{df}{dr}=\sqrt
{E^{2}-1-\gamma^{2}-\frac{L^{2}}{r^{2}}}.
\end{displaymath}
Hence,
\begin{displaymath}\begin{array}{l}
f(r)=L\sqrt
{E^{2}-1-\gamma^{2}-\frac{L^{2}}{r^{2}}}\cr+L\arctan\Big(\frac{1}{\sqrt
{E^{2}-1-\gamma^{2}-\frac{L^{2}}{r^{2}}}}\Big).\end{array}\end{displaymath}
$\psi$ of eq.(30) solves eq.(23) if $r^{2}\geq
L^{2}(E^{2}-1-\gamma^{2})^{-1}\geq 0$. This local solution of
eq.(23) does not define $a(\psi)=\exp(\psi)$ for all $0\leq\phi
\leq 2\pi$ because $\psi(\phi+2\pi)\neq \psi(\phi)$. However, the
local coordinates define the Christoffel symbols (18) depending on
$\partial_{\nu}\psi$ which are functions only on $r$.As a
consequence the Riemann tensor depends solely on $r$ and is
well-defined if $r^{2}\geq L^{2}(E^{2}-1-\gamma^{2})^{-1}\geq 0$.
$v_{\nu}=\partial_{\nu}\psi$ of eq.(30) solve Einstein equations
(1) (which depend only on derivatives of $\psi$). We expect that
there  exists another system of neighborhoods with coordinates
which extend the cylindrical coordinates $(r,\phi,z)$. However,
the range of small $r$ seems to be an essential singularity of the
solution (like $r=0$ of the Schwarzschild solution). The function
(30) describes the solution of the Hamilton-Jacobi equation
corresponding to a relativistic particle moving in the Minkowski
space with a forbidden region of large $L^{2}r^{-2}$
\cite{landau}.
\section{Non-singular solution with the viscous fluid }
 We consider a special case of eq.(30)
  ($L=\gamma=0$, spherical coordinates with $\theta =\frac{\pi}{2}$ and $r=\sqrt{x^{2}+y^{2}+z^{2}}$).
 A spherically invariant solution of eq.(23) which makes sense for arbitrary $(\tau,r)$ can be
expressed as \begin{equation}
\psi=\cosh(\alpha)\tau+\sinh(\alpha)r
\end{equation}
with an arbitrary real $\alpha$.

For a better physical interpretation let us change coordinates
$(\tau,r)\rightarrow (t,R)$ \begin{equation} t=\exp(\tau
\cosh(\alpha)+r\sinh(\alpha))=a
\end{equation}
\begin{equation}
R=\tau\cosh(\alpha)+r\frac{\cosh^{2}(\alpha)}{\sinh(\alpha)}.
\end{equation}
Then, the  metric is
\begin{equation}\begin{array}{l}
ds^{2}=dt^{2}-t^{2}\frac{\cosh^{2}(\alpha)}{\sinh^{2}(\alpha)}dR^{2}
\cr-t^{2}\sinh^{2}(\alpha)(R-\ln(t))^{2}(d\theta^{2}+\sin^{2}\theta
d\phi^{2}). \end{array}\end{equation}
 We can imbed this
universe in the Minkowski space-time ($(\tilde{\tau},\tilde{\bf
x})$ with $ \tilde{\tau}\geq \vert\tilde{\bf x}\vert $)
introducing the coordinates
\begin{equation}
\tilde{\tau}=t\cosh(R\coth(\alpha)),
\end{equation}
\begin{equation}
\tilde{r}=t\sinh(R\coth(\alpha)).
\end{equation}
Then
\begin{equation}\begin{array}{l}
ds^{2}=d\tilde{\tau}^{2}-d\tilde{r}^{2}\cr-\frac{1}{4}(\tilde{\tau}^{2}-\tilde{r}^{2})
\sinh^{2}(\alpha)\ln^{2}D(d\theta^{2}+\sin^{2}\theta
d\phi^{2}),\end{array}
\end{equation}
where
\begin{equation}
D=(\tilde{\tau}-\tilde{r})^{1+q}(\tilde{\tau}+\tilde{r})^{1-q}=
(\tilde{\tau}^{2}-\tilde{r}^{2})\exp(2q\tilde{\eta}),
\end{equation}where we introduced
the space-time rapidity $\tilde{\eta}$( often applied in a
description of heavy ion-collisions \cite{hunperfect})
\begin{displaymath}
\tilde{\eta}=\frac{1}{2}\ln\Big((\tilde{\tau}-\tilde{r})(\tilde{\tau}+\tilde{r})^{-1}\Big)
\end{displaymath}
and
\begin{equation}
q=\tanh(\alpha).
\end{equation}
In these coordinates the radial velocity (the remaining components
of the velocity are zero) has the simple form
\begin{equation}
v_{r}=\sqrt{\tilde{\tau}^{2}-\tilde{r}^{2}}\sinh(\alpha).
\end{equation}
 From eq.(27) the energy density of the fluid is
\begin{equation}
\rho=(\tilde{\tau}^{2}-\tilde{r}^{2})^{-1}\Big(3(1+a\gamma_{g})+4(2+a\gamma_{g})(\ln(D))^{-1}\Big).
\end{equation}(the space-time dependence of $\gamma_{g}$ is not
determined by Einstein equations (1)). The gravitational energy
density can be defined as
\begin{equation}\begin{array}{l}
T^{0}_{0}=G^{0}_{0}=a^{-4}(3(\partial_{0}a)^{2}+(\nabla
a)^{2}-2a\triangle a)\cr
=(\tilde{\tau}^{2}-\tilde{r}^{2})^{-1}\Big(3+2\sinh^{2}(\alpha)+8(\sinh(\alpha)\ln(D))^{-1}\Big)
\end{array}\end{equation} The density of the gravitational momentum (only the component with
radial index is different from zero) is
\begin{equation}\begin{array}{l}
T^{0}_{r}=a^{-4}(4\partial_{0}a\partial_{r}a-2a\partial_{0}\partial_{r}
a)\cr=(\tilde{\tau}^{2}-\tilde{r}^{2})^{-1}4\cosh(\alpha)\sinh(\alpha)\simeq
v_{r}^{-2}\end{array}
\end{equation}

\section{Summary}
We have derived a solution of Einstein equations with a viscous
fluid which is different from the well-known homogeneous solutions
with a perfect fluid.  This kind of  matter could be a constituent
of the models of the universe or could exist in the form of
(expanding) galaxies. The fluid density is decreasing like $a^{-2}
$(if we assume that the bulk viscosity behaves as
$\gamma_{g}\simeq a^{-1}$ like the shear viscosity ) with
logarithmic corrections increasing the decay in comparison to the
coasting cosmology \cite{coasting} . A contribution of such a
fluid to the total energy density (consisting of radiation, dark
matter and "baryons" ) becomes relevant for a large time. It can
be applied in models attempting to explain the coincidence
problem.

\end{document}